\documentstyle[12pt,amssymb,amsfonts,amsbsy]{article}
\begin{document}
\author{Ilja Schmelzer\thanks
       {WIAS Berlin}}

 \title{Postrelativity --- A Paradigm For Quantization With Preferred
Newtonian Frame}

\begin{abstract}
\sloppypar

We define a new paradigm --- postrelativity --- based on the
hypothesis of a preferred hidden Newtonian frame in relativistic
theories.  It leads to a modification of general relativity with ether
interpretation, without topological problems, black hole and big bang
singularities.  Semiclassical theory predicts Hawking radiation with
evaporation before horizon formation.  In quantum gravity there is no
problem of time and topology.  Configuration space and quasiclassical
predictions are different from canonical quantization of general
relativity.  Uncertainty of the light cone or an atomic structure of
the ether may solve ultraviolet problems.  The similar concept for
gauge fields leads to real, physical gauge potential without
Faddeev-Popov ghost fields and Gribov copy problem.

\end{abstract}

\maketitle
\tableofcontents

\section{Introduction}

``Postrelativity'' is a new paradigm about space, time and causality
alternative to the usual, relativistic paradigm of curved spacetime.
The name should suggest that it revives pre-relativistic notions
combined with incorporation of relativistic results.  It is defined by
the following principles:

 \begin{enumerate}

 \item Classical quantum framework: Quantum theory has to be based on
the complete framework of standard, classical quantum theory.

 \item Restricted relativity: Relativistic invariance is not
required. Relativity remains to be a powerful guiding principle, but
only in a restricted sense. A relativistic expression has to be
chosen whenever possible without violation of the first principle.

 \end{enumerate}

The introduction of a new paradigm requires ``postrelativization'' of
current physics --- the development of postrelativistic versions of
existing relativistic theories and reconsideration of relativistic
quantization problems.  The aim of this paper is to give an overview
of the results, hypotheses and ideas we have obtained following this
program.

At first, we give a justification for our choice of principles.
Postrelativity may be understood as a reformulation of the known
``preferred frame hypothesis'' --- the existence of a hidden Newtonian
frame is the most interesting consequence of postrelativity.  This
reinterpretation as an alternative paradigm removes the main argument
against this hypothesis --- incompatibility with the relativistic
paradigm.

Postrelativity already requires a modification of classical general
relativity. We have to incorporate a Newtonian background frame as a
hidden variable. This leads to minor but interesting differences like
a different scenario for the black hole collapse without singularity
and well-defined local energy and momentum conservation. This
postrelativistic theory of gravity suggests an interpretation as an
ether theory with dynamical ether described by deformation tensor,
velocity and a scalar material property --- the local speed of light.
The tetrad formalism may be incorporated, reduces to a triad variant
and suggests interpretation as crystal structure of the ether.

In scalar semiclassical theory we define the configuration space
independent on the gravitational field via canonical quantization.
The vacuum state and the Fock space structure appear only as derived
notions, uniquely defined but dependent on the gravitational field and
time. Hawking radiation is a necessary consequence of this approach.
The black hole evaporates before horizon formation.  The introduction
of tetrad formalism allows to generalize this scheme to particles with
spin.  Using semiclassical considerations the Feynman diagram
technique may be justified only for the first order tree approximation
and for momentum below Planck scale.

In quantum theory, the Newtonian background frame remains fixed and
certain.  On the other hand, the gravitational field, especially the
light cone, becomes uncertain.  The uncertainty of the light cone may
remove ultraviolet problems by regularization of the light cone
singularities.  Another possibility to remove ultraviolet problems may
be the introduction of an atomic structure of the ether, without
necessity of discretization of space or time required by the similar
relativistic concept.

We consider canonical quantization and the path integral formulation
as possibilities for quantization.  We observe some essential
simplifications compared with the canonical quantization of general
relativity.  Especially we have no problem of time and no topological
foam.  We also find a difference in the configuration space.  This
suggests that above approaches lead to quantum theories with different
experimental predictions because of different definition of the Pauli
principle.

For quasiclassical theory we consider a simple gedanken-experiment to
find differences between the postrelativity and relativity.  In
postrelativity we are able to make predictions which coincide in the
non-relativistic limit with Schr\"odinger theory.  These predictions
cannot be made in the relativistic approach because of symmetry
reasons.  This suggests that in the relativistic approach it is
problematic if not impossible to derive Schr\"odinger theory as the
non-relativistic limit.

Then we use the similarity between gauge theory and gravity to find a
version of gauge theory which corresponds to postrelativistic gravity.
In this postrelativistic gauge theory the gauge potential becomes a
real, physical variable, the Lorentz condition has to be interpreted
as a physical evolution equation.  The configuration space of
postrelativistic gauge theory contains different gauge-equivalent
gauge potentials as different states.  This leads to different
experimental predictions at least for non-Abelian gauge theory.
Because of the absence of a gauge-fixing procedure there will be no
Faddeev-Popov ghost fields and no problems with Gribov copies.
Similar to gravity, a quasiclassical gedanken-experiment suggests
problems of the relativistic approach with the Schr\"odinger theory
limit.

Last not least, we discuss some esthetic, metaphysical and
historical questions related with the postrelativistic approach.

\section{The Principles of Postrelativity}

This may be considered as the diametrically opposite to Einstein's
concept that general relativity is more fundamental compared with
quantum theory.

It is known that some problems of relativistic quantum theory,
especially the problem of time \cite{Isham} and the violation of
Bell's inequality \cite{Bell} for realistic hidden variable theories
allow a solution by assumption of a preferred but hidden Newtonian
background frame.

This ``preferred frame hypothesis'' is usually not considered as a
serious alternative.  The reason is that it is not compatible with the
relativistic paradigm --- the philosophical and metaphysical ideas
about space and time related with Einstein's special and general
relativity.  This obvious incompatibility is usually solved in favour
of the relativistic paradigm.  But we consider the problems solved by
the preferred frame hypothesis --- especially the problem of time ---
as serious enough to try the other way and to reject the relativistic
paradigm.  This requires to replace it by an alternative paradigm
which is not in contradiction with the preferred frame hypothesis.

\subsection{A Simple Fictional World}

Let's consider at first a simple fictional world.  This world is
non-relativistic, with a classical Newtonian frame.  By unspecified
reasons, measurement is very restricted, especially for length to
rulers of a single material.  That means, length comparison of
different materials cannot be used to built a thermometer.  We assume
that temperature is not observable by other methods too.

Nonetheless, a non-constant temperature distribution may be observed
by length measurement.  Indeed, it leads to nontrivial curvature of
the metric defined by this length measurement.  On the other hand,
length measurement cannot be used to observe the Newtonian background.
It would be no wonder if it would be able to derive a ``theory of
relativity'' with temperature as an unobservable, hidden potential,
which is able to explain all classical observations.

On the other hand, it is clear that it would not be possible to extend
this relativistic theory to the quantum domain. The correct quantum
theory is --- per construction --- classical quantum theory.  An
identification of states with identical metric but different Newtonian
background would be wrong.

We see, that a situation where the preferred frame hypothesis is
correct and the relativity principle valid in the classical limit but
restricted in general is imaginable.  It may be not an inherent
problem, but only a restriction of our observation possibilities, which
hides the Newtonian background frame.  It would be not the first time
in history we have to learn about the restricted possibilities of
mankind.

In principle, postrelativity may be considered as an attempt to find
out if we live in a similar situation. The two principles we have
formulated for postrelativity can be considered as derived from the
general idea of a hidden preferred frame, by analogy from this
fictional world. Let's now consider these principles in detail.

\subsection{Classical Framework}

At first let's consider the first principle.  It describes the
general, metaphysical and philosophical foundations of the theory and
an essential part of the mathematical apparatus.  It is the apparatus
of classical quantum theory.  This apparatus in no way requires to
reject relativistic field theories.  As we will see below, we don't
have to modify very much to incorporate them into the classical
framework.

Thus, the general structure and the symmetry group of the theory is
classical, relativistic properties follow only from the physics, from
properties of the Schr\"odinger operator.

Of course, this general notion ``framework'' is a little bit uncertain
yet. In some sense, this is natural --- it is possible to modify or
remove some parts from the notion ``framework'' if they cause problems
in future without giving up the whole concept. Thus, the specification
below is also a description of the state of the research which parts
of the classical framework do not cause problems in the following.

\subsubsection{Contemporaneity, Time and Causality}

Absolute contemporaneity is the most important, characteristic part of
the classical framework.  Remark that this contemporaneity is not
considered to be measurable with clocks.  The impossibility of an
exact measurement of time is known from quantum mechanics: Any clock
goes with some probability even back in time \cite{Isham}. Absolute
contemporaneity leads also to absolute causality.

Together with contemporaneity we require symmetry of translations in
time. But, because we have no time measurement, we have no natural
unit. We have only an affine structure in time direction. 

\subsubsection{The General Principles of Quantum Mechanics}

We require the standard mathematical apparatus of quantum mechanics,
that means, the Hilbert space, states as self-adjoint,
positive-definite operators with trace 1, observables as projective
operator measures, and evolution as an unitary transformation.  Any of
the usual interpretations of this apparatus can be used.

We consider classical theory to be only the limit $\hbar \to 0$ of
quantum theory.  In this sense, quantization is an incorrect, inverse
problem, and canonical quantization is considered only as a method to
obtain a good guess.  But, of course, for a given classical theory,
the canonical quantization has to be tried at first.

We do not include into the classical framework any requirements about
the Hamilton operator other than to be self-adjoint and
time-independent.

\subsubsection{Space, Translations and the Affine Galilee Group}

The next required part of classical quantum theory is the
three-dimensional space and the group of translations in space.  This
allows to define position and momentum measurement, the standard
commutation relations and the related standard simplectic structure of
the phase space.

Remark that we have not included a metric of space or time into the
classical framework. Indeed, the metric occurs in the Hamilton
formulation of classical mechanics only in the Hamilton function and
is that's why not part of the framework.  Thus, the symmetry group of
the classical framework as we have defined it here is not the
classical Galilee group, it but the affine variant of this group.
This group contains the following subgroups:

\begin{itemize}
\item translations in space;
\item translations in time;
\item the classical Galilee transformations ($x'^{i} = x^i - v^it$);
\item linear transformations in space ($x^{i'} = a^{i'}_j x^j$);
\item linear transformations in time ($t' = at$);
\item and, of course, any compositions.
\end{itemize}

Remark that because of the absence of a distance in space there is no
preferred subgroup of rotations. In the following we name this group
the ``affine Galilee group''.

\subsubsection{Configuration Space}

The configuration space has to be defined in an affine-invariant way.
Another requirement is locality. That means, field operators dependent
on position for different positions have to be independent.

Another question is the independence between different fields in the
same point. In classical quantum theory, the configuration space is a
tensor product of the configuration spaces for different steps of
freedom.  The interaction is defined by the Hamilton operator, not by
nontrivial configuration space structure.

Such a tensor product structure allows a simple operation --- to
ignore the state of another step of freedom.  We don't have to specify
a complete measurement for all steps of freedom, but can define such a
complete measurement by the measurement of interest for one step of
freedom and ``some other measurement'' for the other steps of freedom.
To ignore other steps of freedom is a measurement of these steps of
freedom which seems easy to realize, thus, it is a natural assumption
that this is possible.

Thus, if for a given theory it is possible to find a formulation with
tensor product structure of the configuration space, it is reasonable
to prefer this variant, at least by Ockham's razor.  Of special
interest is of course the independence between gravity and matter. As
we will see below, it is possible to introduce a tensor product
structure into the configuration space of postrelativistic quantum
gravity.

\subsection{Restricted Relativistic Invariance}

The second principle is formulated in a very weak form, but
nonetheless remains very powerful.  De-facto, all what has been done
in the relativistic domain, with small exceptions, has to be
incorporated because of this principle.  The restriction leads usually
only to one modification: An object which is not relativistic
invariant has in relativity the status ``not existent'', in
postrelativity it has the status ``maybe not observable''.

Roughly speaking, it can be said, that the first principle describes
the framework, the second tells that we have to make the contents as
relativistic as possible. 

\section{Classical Postrelativistic Gravity}

At first, it may be assumed that our set of principles leads to
contradictions already in the classical limit $\hbar \to 0$.  This is
not the case.  We present here a theory which is in full agreement
with the principles of postrelativity and in agreement with experiment
named (classical) postrelativistic gravity (PG).  It can be considered
as a generalization of the Lorentz-Poincare version of special
relativity \cite{Poincare} to general relativity.  It may be
interpreted also as a classical ether theory.  It can also be
considered as a minor modification of general relativity which
de-facto is not distinguishable from general relativity by classical
experiment.

PG can be derived from the postrelativistic principles at least in an
informal way: There have to be an absolute affine time t and an
absolute affine space like in classical theory because of the first
principle. The evolution of the additional variables --- like for all
variables --- has to be fixed by evolution equations.  If possible, we
have to use a relativistic equation because of the relativistic
principle. The existence of such relativistic equations --- the
harmonic coordinate condition is a relativistic wave equation for the
harmonic coordinates --- shows that this is possible and that's why
fixes the harmonic equation as the evolution equation.

\subsection{The Equations of Classical Postrelativistic Gravity}

In the classical formulation we have preferred coordinates - affine
space coordinates and time. The gravitational field is described like
in general relativity by the tensor field
$g_{ij}(x,t)$. General-relativistic time measurement is described like
in general relativity by proper time:

\[\tau=\int\sqrt{g_{ij}(x,t){dx^i\over dt}{dx^j\over dt}}dt\]

The equations of the theory are the Einstein equations

\[R_{ik} - {1\over2}g_{ik}R = {8\pi k \over c^4}T_{ik} \]

and the harmonic conditions

\[ \partial_i \left(g^{ik}\sqrt{-g}\right) = 0 \]

As equations for the material fields we also use the same equations as
general relativity. The coordinates are interpreted as affine
coordinates of a hidden but real Newtonian frame. This defines an
absolute contemporaneity and absolute causality as required by the
first principle. 

Local existence and uniqueness theorems can be easily
obtained. Indeed, the existing results for general relativity use
harmonic coordinates and can be interpreted as theorems for
postrelativistic gravity combined with the derivation of the
general-relativistic results from these theorems.

PG does not define distances for the background structure, nor in
space, nor in time.  Thus, the symmetry group of PG is the affine
Galilee group.

\subsubsection{Covariant Formulation}

Of course, we can give also an equivalent covariant formulation of the
theory.  In this formulation, we introduce a covariant derivative
$\tilde{\nabla}$ (different from the covariant derivative defined by
the metrical tensor) and a global function t and describe them by the
following equations:

\[ \left[ \tilde{\nabla},\tilde{\nabla}] \right] = 0 \]
\[ \tilde{\nabla}_i \left(g^{ik}\sqrt{-g}\right) = 0 \]
\[ \tilde{\nabla}_i \tilde{\nabla}_j t = 0 \]

Let's remark also that the preferred coordinates fulfil relativistic
wave equations:

\[ \Box x^i = 0;\; \Box t = 0; \]

\subsection{Properties of the Theory}

The background framework of PG is hidden from direct observation, but
nonetheless modifies some properties of the theory.  Indeed, the
assumption of additional hidden but real variables can forbids
solutions which do not allow the introduction of these variables, can
modify the definition of completeness of a solution, can change the
symmetry group of the theory and through the Noether theorem the
conservation laws.  All these effects we find in the relation between
PG and GR.  They are similar to the differences found by Logunov
et.al. \cite{Logunov}, \cite{Vlasov} for their ``relativistic theory
of gravity'', a similar theory but with a special-relativistic instead
of Newtonian background.

But, at first, remark that for every solution of PG we can define an
``image''-solution of GR simply by ``forgetting'' the hidden
variables.  That's why, the differences have only one direction: They
allow falsification of PG without falsification of GR.  Let's consider
now the differences in detail. As we will see, PG removes some of the
most complicate problems for quantization: The singularities of the
black hole collapse and the big bang, nontrivial topologies, and the
problems with local energy-momentum tensor.

\subsubsection{Fixation of the Topology}

In a hidden variable theory it may happen that some solutions of the
original theory do not allow the introduction of the hidden variables.
The related observable solutions are forbidden in the hidden variable
theory. This defines one way to falsify the hidden variable theory: To
observe in reality one of the forbidden solutions.

In the case of PG, GR solutions with nontrivial topology are
forbidden. Thus, PG excludes a whole class of GR solutions as
forbidden. Unfortunately, the observation of nontrivial topology, if
it exists, is very nontrivial, because it cannot be restricted to
local observations only. Indeed, it is sufficient to remove some parts
of codimension 1 from the solution with nontrivial topology, and we
have a solution with trivial topology which can be interpreted as a PG
solution, even as a complete PG solution.

Thus, the difference is only of theoretical interest and cannot be
used for a real falsification of PG. Nonetheless, the theoretical
simplification is essential. In quantum PG, we have not to consider
different topologies, thus, we have no topological foam.

\subsubsection{Different Notion of Completeness}

A solution in PG is complete if it is defined on the whole affine
space and time. There is no requirement of completeness of the metric
defined by the gravitational field $g_{ij}$.

The most interesting example of this effect is the black hole collapse.
For a collapsing black hole, there are reasonable initial values of
the harmonic coordinates defined by the Minkowski coordinates in the
limit $t \to -\infty$. The resulting coordinates have the interesting
property that they do not cover the complete GR solution. Indeed, in
the domain outside the collapsing body the harmonic time coincides
with the Schwarzschild time, thus does not cover the part behind the
horizon.

This offers a possibility to falsify PG, which is unfortunately also
only very theoretical. If an observer falls into a black hole created
by collapse, and if he really reaches the part behind the horizon, he
can be sure that PG is falsified. Unfortunately he cannot tell us
about this observation.

Let's remark that the conceptual problems which may be related with
the singularity, especially the possibility that conservation laws may
be violated \cite{Wheeler}, are simply not present in PG. Thus,
another quantum gravity problem has simply disappeared.

\subsubsection{Different Symmetry Group}

Above theories have different symmetry groups. Thus, a solution which
may be considered as symmetrical in GR may not have this symmetry in
PG.  An example are the Friedman universes. Only the flat Friedman
universe allows harmonic coordinates with the same symmetry group.  The
other, curved, solutions allow the introduction of harmonic
coordinates (if we remove a single ``infinite'' world-line from the
closed universe solution), but these solutions are no longer
homogeneous. From point of view of the hidden coordinates, these
solutions have a center.

Thus, PG prefers the flat universe solution as the only homogeneous
one. The fact that the observed universe is at least very close to a
flat universe speaks in favour of PG. But in our world where even P
and CP symmetry is not observed, an inhomogeneous universe is not
forbidden. Thus, observation of a curved universe would not be a
falsification of PG, and that's why it is not possible to say that PG
predicts a flat universe.  Nonetheless, PG suggests an easy
explanation why our universe is approximately flat --- because it is
approximately homogeneous.

 \subsubsection{Local Energy and Momentum Conservation}

Different from GR, we have a preferred symmetry group of translations
in space and time. This leads because of the Noether theorem to
well-defined local conservation laws for energy and momentum of the
gravitational field too. 

This situation is different from general relativity. The
general-relativistic pseudo-tensor $t^{ik}$ is not a tensor, thus,
cannot be observable in general relativity.  This is not required in
PG. Thus, the pseudo-tensor $t^{ik}$ is in PG a well-defined object,
of the same class of reality as the hidden background frame.  Thus,
the problems with the definition of local energy density are not
present in PG. Physically different PG states usually have different
energy even if their general-relativistic image is equivalent.

 \subsection{Triad Formalism}

In general relativity we have some interesting variables known as
tetrad variables. They are useful for the quantization of tensor and
spinor fields on general-relativistic background, and they allow to
replace the non-polynomial $\sqrt{-g}$ by a polynomial expression.
The tetrad variables are four covector fields $e^i_\mu(x,t)$, they
define the metric as

\[g_{\mu\nu}(x,t) = e^i_\mu(x,t)e^j_\nu(x,t)\eta_{ij}\]

In PG, the preferred foliation into space and time already splits the
metric into separate parts. It is natural to require that the tetrad
variables correspond with this splitting. Especially, consider the
hyper-plane defined by constant time.  The time-like tetrad vector can
be defined uniquely by orthogonality to this plane and the direction
of time. Thus, this vector field is already fixed in PG.

The remaining three vector fields now have to be in this plane. Thus,
in PG the tetrad formalism naturally reduces to a triad formalism.  If
we consider these triad variables as the real steps of freedom of the
gravitational field, this leads also to some internal advantages.  The
metric always remains space-like in the plane of constant time.

 \subsection{Ether Interpretation}

PG may be easily interpreted as a dynamical ether theory. At least,
the number of components of the gravitational field $g_{ij}(x)$ is in
good agreement with the steps of freedom and the transformation rules
for a material ether.

Remark that from point of view of PG the gravitational field
$g_{ij}(x)$ splits into parts with separate transformation
behavior. At first, considering the transformation behavior for pure
Galilee transformations, we can identify the vector $v^i =
-g^{0i}/g^{00}$ as the velocity of the ether. The scalar $\rho =
g^{00}\sqrt{-g}$ can be identified as the density of the ether. This
leads to an interpretation of the first harmonic equation as a
conservation law for this density:

\[\partial_t \rho - \partial_i v^i \rho = 0 \]

The space part of the metrical tensor $g_{ij}$ describes the
deformation tensor modulo a scalar factor which defines a scalar
material property --- the local speed of light.

The ether interpretation is in good correspondence with the triad
formalism. The three three triad vector fields define some hidden
preferred directions, which suggest an interpretation in terms of a
crystal structure of the ether.

The ether interpretation can also give hints for quantization. For
example, if the harmonic equation is a conservation law, it has to be
fulfilled also for quantum configurations, thus, interpreted as a
constraint, not an evolution equation.

But much more interesting is that it suggests that there may be an
underlying atomic structure of the ether.  Such an atomic variant
highly probable allows to solve ultraviolet problems.

\section{Semiclassical Postrelativity}

Semiclassical theory considers quantum fields on a fixed, classical
background solution for the gravitational field.  Thus, we consider
the quantum effects only as small and neglect the influence of the
quantum effects on the gravitational field. 

\subsection{Canonical Quantization of the Scalar Field}

Assume we have given a PG solution as the fixed background. Consider a
scalar particle on this background with the standard relativistic
Lagrangian (Greek indices from 0 to 3, Latin indices from 1 to 3):

\[\label{defLscal} 
{\cal L} = 
{1\over2}\sqrt{-g}(g^{\mu\nu}\phi_{,\mu}\phi_{,\nu}-m^2\phi^2) \]

Using the standard canonical formalism, we define 
($\label{defhatg}\hat{g}^{\mu\nu}={g}^{\mu\nu}\sqrt{-g}$) 

\[\label{defpiscal} 
\pi = {\partial{\cal L}\over\partial\phi_{,0}} 
= \hat{g}^{0\mu}\phi_{,\mu} \]

\[\label{defHscal} 
{\cal H}=\pi\phi_{,0}-{\cal L}
={1\over2} (\hat{g}^{00})^{-1} (\pi-\hat{g}^{0i}\phi_{,i})^2
-{1\over2}\hat{g}^{ij}\phi_{,i}\phi_{,j}
+ {m^2\over2}\phi^2\sqrt{-g} 
\]

We define now $\phi$ and $\pi$ as operators with the standard
commutation rules ($\hbar=1$):

\[\label{defcancomm}
  [\phi(x), \pi(y) ] = i\delta(x-y)
\]

This already gives the guarantee that the first principle is
fulfilled. We don't consider here the problem of ordering which occurs
in the definition of the Schr\"odinger operator.  We define the vacuum
state as the state with minimal energy, which has to exist because the
energy is nonnegative.

\subsection{Special-Relativistic Quantum Field Theory}

In the case of the Minkowski space $\hat{g}^{\mu\nu}=\eta^{\mu\nu}$ it
is useful to introduce another basis:

\[ \phi_k = \int e^{ikx}\phi(x)dx,\; \pi_k = \int e^{ikx}\pi(x)dx \]

which essentially simplifies the expression for the energy

\[H = \int {\cal H} dx = {1\over2} \int \pi_k^2 
+ \omega_k^2 \phi_k^2
dk, \omega_k = \sqrt{k^2+m^2} \]

Now it is useful to define operators

\[\label{defapmMinkowski}
a^{\dag}_k={1\over\sqrt{2\omega_k}}(\pi_k+i\omega_k\phi_k)\;
a_k       ={1\over\sqrt{2\omega_k}}(\pi_k-i\omega_k\phi_k)
\]

with the well-known commutation relations with H:

\[ [a_k,H] = \omega_k a_k \]

These operators allow to characterize the vacuum state, because for
this state we have

\[ a_k |0\rangle = 0 \]

Any other constant gravitational field (a metric of the same signature
as for the whole spacetime, as for the space only), can be transformed
in PG to the standard Minkowski form within the PG symmetry
group. Nonetheless, let's write down the related expression for the
general constant field too:

\[a^{\dag}_k = {1\over\sqrt{2\omega_k}}(\pi_k -i(\hat{g}^{0i} k_i 
- \omega_k)\phi_k),\]
\[a_k        = {1\over\sqrt{2\omega_k}}(\pi_k -i(\hat{g}^{0i} k_i
+ \omega_k)\phi_k),\]
\[\omega_k^2     = \hat{g}^{00} (-\hat{g}^{ij}k_ik_j+m^2\sqrt{-g})\]

\subsection{Nontrivial Gravitational Field}

In the general case, it is not so easy to define such a
basis. Nonetheless, let's try to characterize the vacuum state in a
similar way at least approximately. Indeed, consider the wave packets

\[ 
\phi_{kx} = \int e^{iky-\sigma(y-x)^2}\phi(y)dy,\;
\pi_{kx}  = \int e^{iky-\sigma(y-x)^2}\pi (y)dy;
\]

Assume a sufficiently small $\sigma$, so that we have a good
approximation of the momentum, but, on the other hand, assume $\sigma$
big enough so that the gravitational field can be approximated by a
constant field in the region there the function is not very small.  In
this approximation, we can define local particle operators

\[a^{\dag}_{kx}={1\over\sqrt{2\omega_{kx}}}
(\pi_{kx}-i(\hat{g}^{0i} k_i-\omega_{kx})\phi_{kx}),\]

\[a_{kx}       ={1\over\sqrt{2\omega_{kx}}}
(\pi_{kx}-i(\hat{g}^{0i} k_i+\omega_{kx})\phi_{kx}),\]

\[\omega_{kx}^2 =\hat{g}^{00} (-\hat{g}^{ij}k_ik_j+m^2\sqrt{-g});\]

We obtain

\[ [a_{kx},H] \approx \omega_{kx} a_{kx} \]
\[ a_{kx} |0\rangle \approx 0 \]

Thus, for gravitational fields which vary not very fast, the vacuum
state looks locally like the Minkowski vacuum.

\subsubsection{Hawking Radiation}

In this way, semiclassical PG defines a natural vacuum state and Fock
space structure dependent on the gravitational field and time.  In
general relativity we have no such natural Fock space definition.  In
the GR paradigm, this is explained in the following way: To define the
notion of a particle, we have to choose a preferred set of particle
detectors which are considered to be inertial. The vacuum would be the
state where these particle detectors do not detect particles.

The previous considerations allow us to describe the vacuum state
defined by PG in these words too. Indeed, the PG background structure
defines a preferred set of observers which are considered as inertial
observers. At a given moment of time, these local observers do not
observe particles in the vacuum state.

It is essential that the vacuum state definition in PG depends on
time.  The state which is the vacuum at $t=t_0$ in general not become
the vacuum state at $t=t_1$, but becomes a state with particles.  This
effect leads to Hawking radiation.  Let's show this: Outside the
collapsing body, the harmonic coordinates which initially coincide
with the Minkowski coordinates are static, especially the harmonic
time is simply Schwarzschild time.  As we have shown, in the vacuum
state at least far away from the surface an observer at rest does not
detect any particles. Thus, this vacuum state will be close to the
state known as Boulware vacuum state. The real state is of course the
result of the evolution of the initial Minkowski vacuum.  The
evolution of this state - known as the Unruh state - coincides after
the collapse with the Hartle-Hawking state. In this state, observers
at rest relative to the black hole observe Hawking radiation.

Thus, Hawking radiation is a natural consequence of postrelativity for
the Fock space definition described here. The conceptual problems with
the uncertainty of the definition of the Fock space do not occur in
postrelativity.

\subsubsection{Scenario of Black Hole Evaporation}

Because we have no horizon formation in classical PG, we already know
that Hawking radiation starts before the horizon is formed.  If we
consider the classical background as fixed, we observe a small but
constant loss of energy in Schwarzschild time.  Because of energy
conservation this has to be compensated by a modification of the
energy-momentum tensor of the classical background solution at this
Schwarzschild time, that means, before horizon formation.

This modification leads to a decreasing horizon, thus, if the
unmodified surface hasn't reached the horizon, the modified hasn't
reached it too.  Thus, this modification does not lead to a
modification of the property that the horizon will not be reached by
the surface in PG.  Combined with the known results about the time for
evaporation for the outside observer, which is finite, it seems clear
that the black hole evaporates in PG before the horizon is even
formed. Of course, there may be modifications of this picture near
Planck length, for example there may be a remaining black hole of
Planck order size without Hawking radiation. Nonetheless, even in this
state we have no singularity inside.

The answer of general relativity about the evaporation is not so
certain. According to Birrell and Davies \cite{Birrell}, there are
different proposals, with exploding singularity or remaining
singularity of Planck order, but also a similar proposal of
evaporation before horizon formation \cite{Gerlach}.

\subsection{Nonscalar Fields}

\subsubsection{The Dirac Field}

Before considering the Dirac field, remark that we have been able to
define the configuration space for the scalar field independent of the
gravitational field.  Such a property is suggested by our first
principle, thus, it seems natural to try to get the same independence
for the other fields too.

Let's try to do this for the Dirac field. In special-relativistic
field theory, we have the definition

\[\left\{\psi^+_\alpha(x),\psi_\beta(y)\right\}_+ 
= \delta_{\alpha\beta} \delta(x-y) \]

which depends only on the metric $\delta_{\alpha\beta}$, but not on
the Minkowski metric.  If we try to use this as the definition of the
configuration space, we immediately obtain a problem: We have to
establish a relation between the operators $\gamma^i$ and the partial
derivatives $\partial_i$. If we fix such a relation, it defines a
Minkowski metric in the space derived from the internal Minkowski
metric. Thus, this relation cannot be independent from the
gravitational field. That means, this relation cannot be part of a
gravity-independent configuration space for the Dirac field.

Thus, it has to be part of the gravitational field.  That means, the
gravitational field has to define an isometric relation between a
four-dimensional internal Minkowski space defined by the $\gamma^i$
and the tangential Minkowski space. The metric $g_{ij}(x)$ alone does
not allow to define such a relation.

This problem is solved by the tetrad formalism.  Thus, to be able to
describe the Dirac field with an independent configuration space, we
have to introduce tetrad variables into postrelativistic gravity.
After the introduction of tetrad variables we are able to define the
configuration space of the Dirac field by the same definition as for
the Minkowski space.

\subsubsection{Other Fields and Interactions}

The tetrad technique can be used for other spinor and tensor fields
too.  The problem is reduced in this way to the definition of the
configuration space for a standard Minkowski frame.  Gauge fields we
consider separately below.

The introduction of interaction terms changes only the Schr\"odinger
operator, thus, does not have any influence on the configuration
space.  Thus, nothing has to be changed compared with the situation
for free fields.  That means, it is possible to derive the Feynman
diagram formalism. 

But it has to be remarked that the semiclassical limit is only
justified for momentum below Planck scale. Already the Fock space is
defined only in this sense.  Thus, it is not justified to consider any
integral over the whole momentum space. That means, only the first,
tree approximation is justified.  Thus, it is clear that at Planck
scale the semiclassical limit becomes wrong.

Thus, we have no reason to wonder about ultraviolet problems in such
illegal integrals.  Of course, on the other hand, the results obtained
by renormalization are reasonable.  It is reasonable to assume that
the correct theory leads to finite terms based on some type of
effective cutoff at the order of Planck scale with unknown details.
Renormalization claims to be able to compute results which do not
depend on these details, even on the order of magnitude of the cutoff.

\subsubsection{Small Quantum Variations of the Gravitational Field}

The semiclassical approximation may be applied to consider small
modifications of the gravitational field too. This leads to a standard
Feynman diagram scheme for general relativity in harmonic gauge.

The independence of the configuration spaces of matter from gravity is
necessary to show the correctness of the consideration of small
modifications of the gravitational field as an independent quantum
field. Else, the consideration of any material field operator together
with operators which describe the difference between the real
gravitational field and the background metric would be meaningless.

Especially that means that this approach is meaningless in the context
of general relativity.  The difference between even very close
solutions of the Einstein equation is not covariant, thus not
defined. Applying different coordinate transformations for above
fields we can make the difference arbitrary big and strange. This
difference between the relativistic and the postrelativistic approach
if different gravitational fields are involved is discussed below.

\section{Ultraviolet Problems and Non-Renormalizability}

It is known that quantum general relativity is non-renormalizable. We
havn't modified anything which may change this fact, thus,
postrelativistic gravity is non-renormalizable too.  Many people
consider this as a strong, even decisive argument against a theory.
There are already arguments \cite{Ashtekar} which show that
non-renormalizability it is not a decisive argument against a theory.
But, on the other hand, it is considered as a serious conceptual
problem.

In the context of postrelativity I consider non-renormalizability not
as a conceptual, but only as a technical problem.  Some of the
following remarks to justify this are valid also for the relativistic
approach, other not.

First, we have already seen that there is no reason to wonder about
infinities.  They simply show an obvious error --- the attempt to
apply the semiclassical limit outside it's possibilities.  In this
sense, the divergences of these integrals not a conceptual problem,
because our concepts don't even suggest these integrals have to be
finite.

They are also not a problem to justify the first order tree
approximation.  Our concepts also don't suggest a definition of
quantum gravity via a formal power series based on a fixed classical
background.  We have derived the semiclassical limit based on some
general assumptions about the correspondence between the unknown full
theory and their semiclassical limit.  Thus, the derivation of the
first order tree approximation is based on these assumptions, not on
the formal power series and the correct definition of the higher order
terms.  Thus, problems of computation of higher order approximations
do not question the correctness of the first order tree approximation.

\subsection{Light Cone Uncertainty}

Second, let's consider a simple qualitative prediction about the
properties of postrelativistic quantum gravity.  This prediction is
the uncertainty of the light cone. 

Remark at first that such a prediction does not make sense from point
of view of the relativistic paradigm, because this paradigm does not
allow to compare different solutions. This property of the
relativistic approach we consider in detail below.  But in the
postrelativistic approach we can compare the light cone of different
solutions.  From point of view of the postrelativistic paradigm,
events are defined independent of the state of the gravitational
field.  It makes sense to compare different gravitational fields. And
we observe easily that different gravitational fields usually have
different light cones. That means, in quantum PG the light cone will
be uncertain.

There is no possibility to avoid this effect in postrelativity.  But
we have also no reason to bother, because this uncertainty does not
cause any problem.  It is not dangerous nor for causality, nor for
position, because above notions are defined independent of the state
of the gravitational field.

Moreover, this uncertainty suggests that ultraviolet problems of the
usual type are not present in quantum PG. Indeed, the ultraviolet
problems in relativistic quantum theory are caused by the singularity
of the propagator near the light cone.  But, if the light cone is not
defined exactly, where is no place left for a light cone singularity
to survive.

This is in no way a proof of anything. But it nonetheless suggests
that a correct computation of higher order approximation (different
from the incorrect one which remains in the semiclassical
approximation without justification) may not lead to infinities.  Last
not least, we have a new physical effect --- the uncertainty of the
light cone caused by the quantum character of the gravitational field.

\subsection{The Atomic Ether}

Third, the ether interpretation of PG suggest a simple way to avoid
ultraviolet problems --- the assumption of an atomic ether.  Indeed,
to make this assumption is even natural without having any ultraviolet
problem, because of the same philosophical reasons which have
justified the atomic hypothesis for usual matter.  

If we make such an assumption, we obviously obtain an effective cutoff
which depends on the typical distance between the atoms of the ether.
Thus, the assumption of an atomic ether defines a simple emergency
exit for the case that the previous ideas do not lead to a removal of
the ultraviolet problems.

The idea to introduce a discrete structure to solve the ultraviolet
problems is not new \cite{Ashtekar}.  But the realization of this idea
in the relativistic paradigm leads to a completely different concept
--- a discrete spacetime.  It requires completely different
mathematics and foundations.  Compared with this idea, the atomic
ether is a very simple idea.  Of course, we may obtain a lot of
difficult technical problems, but conceptually the atomic ether is as
complicate as a deformed crystal material.

Nonetheless, it seems not yet the time to develop such a theory in
detail, this would be too speculative.  

\subsection{The Status of the Ultraviolet Problem}

Let's remark that the status of ultraviolet problems is different in
relativistic and postrelativistic theory.  I don't want to diminish
the technical problems, but I reject to consider ultraviolet problems
as a serious {\it conceptual} problem of the postrelativistic
approach.

Indeed, GR claims to be a theory of space and time, that means about
the most fundamental things in the universe. It claims to be able to
predict the evolution even of the topology of our space. There is
nothing more fundamental than space and time. Thus, an ultraviolet
problem becomes a serious conceptual problem in our understanding of
space and time if they occur in this theory.

The status is completely different in postrelativity. PG doesn't claim
to be the ultimate theory about space and time, it is a continuous
ether theory, with similarity to classical continuum mechanics.  If
ultraviolet problems occur in such a continuous ether theory, they
only show that the ether has some different, probably atomic,
microscopic structure which is not yet observable.  Thus, as far as we
do not pretend to have found the ultimate theory --- which is not a
very natural claim for a continuous ether theory --- this does not
even suggest that there is anything wrong with our continuous
approximation, and is that's why not a conceptual flaw of this theory,
but the only chance for future research to observe --- at least in
principle --- the microscopic structure of the ether.

\section{Canonical Quantization and Path Integral Formulation}

We have seen that there is no reason to be afraid of ultraviolet
problems --- they do not occur immediately in the semiclassical
approximation, there are reasons to suggest that they do not occur in
higher order approximations too, and we have an emergency exit if they
nonetheless occur.

Nonetheless, perturbative theory starting with a classical background
solution does not suggest a way of rigorous definition of quantum PG.
For this purpose, other methods have to be used.

Two methods may be considered for this purpose --- Feynman path
integral formulation and canonical quantization.  For above concepts,
postrelativity leaves some freedom. Nonetheless, we can compare our
approach with the standard general-relativistic paradigm. We can show
not only a difference, but also an essential technical and conceptual
simplification.

\subsection{Freedom of Choice for Further Quantization}

For above methods our first principles leave some freedom of choice
for the following steps.  Indeed, above concepts require to fix the
following:

\begin{itemize}
\item the configuration space and
\item the Lagrange function.
\end{itemize}

The classical postrelativistic theory leaves here some freedom. We
have different choices which lead to the same classical equations:

 \begin{itemize}

 \item We can consider the harmonic equation as an external
constraint. In this case, only harmonic fields are valid field
configurations. This would be natural if we interpret it as a
conservation law.

 \item The other alternative would be to consider it as a classical
evolution equation, and to add a penalty term to the Lagrange
functional so that the Euler-Lagrange equations include not only the
Einstein equations, but also the harmonic equation. In this case, all
field configurations (inclusive non-harmonic) are valid. The gauge
field correspondence considered below suggests this choice.

 \item Orthogonal to this question, we have the possibility to
introduce other variables with another configuration space.  We have
already seen that the introduction of the triad formalism is
reasonable.  This formalism introduces some new hidden (gauge) steps
of freedom.

 \item But, of course, also other variables like Ashtekar's variables
\cite{Ashtekar} have to be considered.

 \item We also have some freedom how to choose the Lagrange function
for gravity between the usual Lagrange density --- the scalar R ---
and the Rosen Lagrangian. The first is covariant, the second not. On
the other hand, the first includes second derivatives, the second
not. Postrelativity suggests to use the second, because second
derivatives will be a problem, non-covariance not.

 \item If we introduce delta-functions into the path integral, we have
to bother about correct norm. Thus, we may have to include an
appropriate normalization coefficient.

 \end{itemize}

Because of the Pauli principle, at least different choices of the
configuration space lead with high probability to different quantum
theories. Thus, postrelativity does not fix quantum gravity uniquely.
To find which is the best choice has been left to future research.  My
personal preference at the current moment is the Rosen Lagrange
function, harmonic equation as a constraint, triad variables.  But one
in in no way forced in this direction.


\subsection{Properties of Quantum Postrelativistic Gravity}

Nonetheless, all these variants have common properties, which we will
describe here as properties of quantum postrelativity. 

 \begin{itemize}

 \item Postrelativistic gravity is the classical limit.

 \item The path integral is defined between arbitrary but fixed, finite
moments of time $t_0, t_1$.

 \item The configuration space consists of functions defined on the
three-dimensional affine space. Especially the functions $g_{ij}(x)$
are defined and describe the gravitational field.

 \item Configurations with different gravitational field $g_{ij}(x)$
are different, even if the configurations can be transformed into each
other by diffeomorphism. For such configurations, the related
probabilities have to be added, not the amplitudes.

 \item For canonical quantization, we obtain a well-defined evolution
in time, different from the Wheeler-DeWitt equation in canonical
quantization of general relativity.

 \end{itemize}

\subsection{Comparison With Relativistic Approach}

On the classical level, the main difference between the relativistic
approach and the postrelativistic approach is the consideration of
configurations which can be transformed into each other with a
diffeomorphism. In GR they are identified, in PG they are different.

This identification leads to conceptual problems even in the
formulation of quantization. The first and most serious group of
problems is connected with diffeomorphisms which change time. The
related problems are known as the ``problem of time''.  Only if we
neglect this problem by considering only diffeomorphisms which don't
change time, we are able to define a configuration space which may be
compared with the PG configuration space. This comparison shows that
the configuration spaces are essentially different. Thus, highly
probable, the resulting quantum theories will be different too, simply
because of the Pauli principle.

\subsubsection{Problem of Time}

According to the paradigm, configurations have to be identified if
there is a diffeomorphism between the configurations.

In this sense, it is already a violation of the paradigm if we write
down a path integral with finite, fixed boundaries for time $t_0$,
$t_1$.  Probably only path integrals with infinite limit or over
compact solutions can fulfil the paradigm completely.

This occurs in the canonical quantization approach too. As the result,
instead of a Hamiltonian evolution we obtain only a so-called Hamilton
constraint. After quantization, this leads to the Wheeler-DeWitt
equation

\[ \hat{H} \psi = 0 \]

instead of a usual Schr\"odinger equation. This equation is considered
to describe only the diffeomorphism-invariant information about our
world. Thus, similar to the problem in the path-integral formulation,
we have no description of the evolution for any finite time, how to
extract physically meaningful information is completely unclear.

Because we are not able to solve this problem, we ignore it and
consider in the following only diffeomorphisms which leave the time
coordinate invariant.

\subsubsection{Topological Foam}

A second problem is the topology of the space. The topology is usually
not a problem in classical general relativity, because it is
controlled more or less by the Einstein equation which usually does
not change topology during the evolution. But in the quantum domain,
we have to consider also non-classical configurations. Because for
small distances we have to assume that quantum theory allows small
variations of the field, inclusive small, local variations of the
topology, this concept leads to the picture that at small distances
the topology will be de-facto uncertain. This concept is known as
``topological foam''.

Because we are not able to solve this problem, we fix in the following
the topology of the space, for simplicity we consider only trivial
topology.

\subsubsection{The Configuration Space of General Relativity}

After these two simplifications we can at least define the
configuration space of the general-relativistic approach in a form
comparable with the postrelativistic configuration space. It is the
result of factorization of the postrelativistic configuration space
where diffeomorph configurations have been identified.

Thus, we see, that two essential simplifications have been necessary
even to define a configuration space which may be compared with
postrelativity.  Moreover, the resulting configuration space is
essentially different.  This highly probable leads to a quantum theory
with different experimental predictions, simply because in the path
integral we have a different basic rule for the computation of
probabilities. Indeed, if different but diffeomorph configurations are
considered, we have to add the related probabilities in the
postrelativistic approach, but the related amplitudes in the
general-relativistic approach. Thus, already the Pauli principle is
defined differently.  Probably in some situation one theory will
predict interference effects but the other not.

\section{Quasiclassical Theory}

In this section we compare the predictions of above theories in the
quasiclassical situation. That means, we leave the semiclassical
situation where the gravitational field is approximated by a classical
field and consider the next step --- superpositions of such states.

In this case, above concepts lead to different predictions.  More
accurate, the general-relativistic concept remains silent, we are not
able to obtain predictions from this concept. Nonetheless, the
prediction of postrelativity cannot be copied, because it is in
contradiction with the symmetry principles of the approach. Thus, it
is reasonable to say nonetheless that the predictions are different.

\subsection{Non-Relativistic Description of a Simple Experiment}

At first, let's describe our experiment in non-relativistic language.
More accurate, we describe it using classical multi-particle
Schr\"odinger theory with Newtonian interaction potential.

We consider a ``heavy'' particle in a simple superposition state

\[ \psi = \delta(x-x_1)+\delta(x-x_2) \]

and it's gravitational interaction with a light test particle. The
initial product state splits into a nontrivial two-particle state. To
compute this state, we use quasiclassical approximation, thus, if the
heavy particle is in the delta-state, we approximate the two-particle
problem by a single-particle problem for the test particle in the
classical gravitational field created by the heavy particle in this
position:

\[ i\partial_t \phi^{1/2} = H^{1/2} \phi^{1/2} \]

\[ H^{1/2} = {p^2\over 2m} - {k\over |x-x_{1/2}|} \]

Then we interpret this one-particle solutions as two-particle tensor
product states and use standard superposition rules to compute the
result:

\[ \phi\otimes\psi \to \phi_1\otimes\delta(x-x_1)
+\phi_2\otimes\delta(x-x_2) \]

After the interaction, we simply ignore the test particle, but
measure, if the state of the heavy particle has changed or not. This
is simple and can be done by an arbitrary interference experiment
which tests if the state of the heavy particle is yet in a
superposition state or not, or, in other words, if the interaction
with the test particle was strong enough to be considered as a
measurement of the position or not. The probability of observing the
heavy particle unchanged is

\[\rho = {1\over2} (1 + {\cal R}e \langle \phi_1 |\phi_2\rangle) \]

The extremal case of scalar product 1 can be interpreted as no
measurement by the interaction with the test particle, thus we observe
interference, and the other extremal case of scalar product 0 as
complete position measurement, thus we observe no more interference.

But the point of this consideration is that the real part of the
scalar product is observable in non-relativistic Schr\"odinger theory.

\subsection{Postrelativistic Description of the Experiment}

Let's describe now the same experiment in the postrelativistic
approach.  In principle, we can use a similar, classical language.  We
need only small modifications. Instead of the one-particle
Schr\"odinger equation with the operator $H^{1/2}$ we have to consider
now the semiclassical theory for fixed classical background
$g_{ij}^{1/2}(x,t)$.  The functions $\phi^{1/2}(x,t)$ are replaced by
states defined in the configuration space of semiclassical theory.

It becomes essential now that the definition of the configuration
space itself was given in terms of the operators $\phi$ and $\pi$
independent of the gravitational field, not in terms of the particle
operators which depend on the field. Thus, the two states
$|\phi^{1/2}\rangle$ are states in the same Hilbert space. Thus, we
can define their scalar product without problem.

Thus, the postrelativistic approach makes clear predictions about the
results of the experiment. It allows to compute the relativistic
corrections.  These predictions coincide in the non-relativistic limit
with classical Schr\"odinger theory.

\subsection{Non-Covariance of the Scalar Product}

Consider now the situation in general relativity. Let's use the
language introduced by Anandan \cite{Anandan} who has considered a
similar superposition experiment. If there is a superposition of
gravitational fields, he distinguishes two types of diffeomorphism: a
classical or c-diffeomorphism that is the same for all superposed
gravitational fields, and a quantum or q-diffeomorphism which may be
different for the different superposed fields. He postulates as the
``principle of quantum general covariance'' that all physical effects
should be invariant under all q-diffeomorphisms.

As we can easily see, the scalar product cannot be observable in this
approach.  Indeed, the semiclassical theory allows to define the
states $|\phi^{1/2}\rangle\otimes|g^{1/2}\rangle$ only as pairs
$(\phi^{1/2}(x,t),g_{ij}^{1/2}(x,t))$ modulo arbitrary coordinate
transformations $(x,t) \to (x',t')$:

\[(\phi^{1/2}(x,t),g_{ij}^{1/2}(x,t)) 
\to (\phi^{1/2}(x',t'),g_{ij}^{1/2}(x',t')) \]

The scalar product as defined in postrelativity

\[ \int \phi^1(x,t) \bar{\phi}^2(x,t) dx \]

is obviously invariant only for c-diffeomorphisms, but not for
q-diffeomorphisms.

Anandan's principle is a consequence of the Einstein equations and
cannot be simply removed from quantum general relativity.  Indeed, if
we consider a superposition of semiclassical solutions, above
solutions are only defined modulo an arbitrary diffeomorphism, thus,
quasiclassical general relativity is automatically
q-diffeomorphism-invariant, if we don't introduce some new
non-q-diffeomorphism-invariant mechanism into the theory.  Moreover,
the configuration space and the path integral formulation which we
have considered for the general-relativistic approach also requires
that the resulting quantum theory is q-diffeomorphism-invariant.

Thus, the scalar product is not defined in the general-relativistic
approach, observable results of this theory cannot depend on such
scalar products.  That means, we are not able to predict
relativistic corrections of our simple experiment. 

\subsection{Remarks About the Seriousness of this Problem}

In principle, this problem can be added to the list of already
existing conceptual problems of the general-relativistic approach
which do not occur in the postrelativistic approach, but nonetheless
continue to hope for a solution of all these problems in future. But
in my opinion it has to be considered as a decisive argument in favour
of the postrelativistic approach. Some remarks in favour of this
position:

Remark that it is not our personal inability to compute the
predictions of the general-relativistic approach, but a clear symmetry
requirement of general relativity which does not allow to define this
scalar product.

Remark that the argument is present in full beauty in the classical
limit.  Indeed, we can rewrite the classical Schr\"odinger theory
experiment in general-relativistic language, using the metric

\[ g_{00} = 1 + {2\phi\over c^2} \]

and restricting the consideration to small velocities. We already have
different gravitational fields, thus, the full problem of
q-diffeomorphism-invariance. That this is not an exact solution of the
Einstein equations is not significant, because in quantum theory we
have to be able to handle configurations which are not exact
solutions.  This suggests that a q-diffeomorphism-invariant theory
will not have Schr\"odinger theory as the classical limit.

Remark that the problem is present already for very small
modifications of the gravitational field. 

Remark that if we are able to define scalar products between functions
on different solutions, we have de-facto a diffeomorphism between the
solutions.  Indeed, we can simply consider the scalar products between
delta-like functions.  And having a diffeomorphism for any two
solutions is very close to a coordinate condition. Indeed, a
diffeomorphism between the Minkowski space and an arbitrary space
defines a preferred coordinate system --- affine Minkowski coordinates
--- on the other solution.

Remark that the idea to accept a break of the
q-diffeomorphism-invariance temporary, as a gauge condition, does not
help if we want to obtain the observable prediction of classical
Schr\"odinger theory.

Remark that the idea to accept a break of the
q-diffeomorphism-invariance but to make it as relativistic as possible
is de-facto the postrelativistic approach. Indeed, we use a really
beautiful relativistic wave equation to define the scalar product.  In
this sense, the postrelativistic approach can be considered as the
simplest way to solve this problem.

Remark, that we have simply ignored the results of measurement of the
test particle.  The aim of this ignorance was to avoid the
consideration of problems which are related with almost every
measurement in the relativistic approach.  In the postrelativistic
approach, we are able to consider the results of some measurement, for
example coordinate measurement, for the test particle too.

On the other hand, the result depends on assumptions about
measurability in Schr\"odinger theory.  It may be argued that this
theory may be wrong or that these observables are not really
observable, but observable only in the classical limit.  

Such argumentation may be of course used to show that nonetheless this
problem is not decisive.  But, of course, a theory of quantum gravity
has to be based on some assumptions which cannot be exactly proven.
This leads to the question how a more decisive argument against the
relativistic approach could look like.

\section{Postrelativistic Gauge Theory}

The postrelativistic principles do not define immediately what has to
be done with gauge fields.  But there is a close similarity between
gauge theory and general relativity. This suggests that there has to
be also a similar correspondence between gravity and gauge theory in
postrelativity too.

Using this correspondence argument, we obtain a new approach for gauge
theory.  It seems natural to use the name ``postrelativistic gauge
theory'' for this gauge-theoretical approach too.  Nonetheless, it has
to be recognized that postrelativistic gravity and postrelativistic
gauge theory are different, independent theories.  Failure or success
of one of them does not immediately prove failure or success of the
other.  But, of course, the correspondence will be a strong
correspondence argument.

The main property of postrelativistic gauge theory is that the gauge
potential has to be considered as a hidden but real step of freedom.
The Lorentz condition becomes a physical equation, not a gauge
condition.  

Classical postrelativistic gauge theory cannot be distinguished from
the relativistic variant. In the quantum domain, they become
different.

 \subsection{Standard Paradigm --- No Gauge Freedom}

Let's shortly remember the standard, usual paradigm, which corresponds
to general relativity.  In this paradigm, different but
gauge-equivalent gauge potentials are identical. The gauge freedom is
considered only as a mathematical construct which makes it easier to
write down some formulas, not as a real freedom.  In the path
integral, the usual way to realize this is to use a gauge condition
which defines a unique gauge potential for each class of
gauge-equivalent potentials:

\[\label{pathintgauge}
\int_{t_0}^{t^1} \exp i\int {\cal L} dt 
\prod_{x,t}\Delta({\cal A}) \delta(f({\cal A})) d{\cal A}
\]

Every equivalence class has to occur in the integral only once.  It
would be even more beautiful if we could describe gauge fields
immediately in gauge-invariant terms like Wilson loops.

The most interesting (because of their relativistic form) gauge
condition --- the Lorentz condition --- solves only half of the
problem of gauge fixing. Indeed, it fixes the gauge only for fixed
boundary conditions, but doesn't fix the gauge for the boundary
conditions.  This remaining gauge freedom has to be fixed by other,
additional boundary conditions.  This type of gauge fixing leads to
problems with unitarity of the S-matrix, if it is not compensated by
additional terms. These compensation terms may be interpreted as terms
describing particles known as Faddeev-Popov ghost particles. But in
the general case even fixed boundary conditions may be not sufficient
to fix the gauge with the Lorentz condition --- there may be so called
Gribov copies. The problem is that the Gribov copies occur in the path
integral as different states, but have to be identical in the ideal
theory.

 \subsection{Classical Postrelativistic Gauge Theory}

The general correspondence between gauge theory and general relativity
requires to consider the gauge freedom as the analog of the freedom of
choice of coordinates in general relativity, gauge transformations as
the analog of diffeomorphisms.  The definition of the gravitational
field in a given coordinate system is the analog of the gauge
potential. Let's apply this correspondence scheme to the
postrelativistic approach.

The configuration space of postrelativistic gravity consists of
gravitational fields in given coordinates.  The analog of this
configuration space is obviously the space of all gauge
potentials. Thus, in postrelativistic gauge theory the gauge potential
is the real variable we have to use to describe the gauge field. Field
configurations which are different but equivalent from point of view
of the symmetry transformation of the relativistic theory are
considered as different states in the postrelativistic theory.

Thus, as in postrelativistic gravity, we have to introduce new steps
of freedom into the theory. They are not directly observable.  To
describe the evolution of these observables, we need a new equation. 

For gravity we have used an equation known already as a very useful
coordinate condition, the harmonic condition. The similarity to the
Lorentz condition in gauge theory is obvious: Above conditions can be
written as a first-order divergence-like condition for the variables
we use to describe the fields, but also as a second order relativistic
wave equation immediately for the hidden steps of freedom. That's why
in postrelativistic gauge theory we consider the Lorentz condition as
a physical evolution equation for the gauge potential.

 \subsection{Canonical Quantization}

As suggested by general rules, let's try now canonical quantization of
classical postrelativistic gauge theory.  We have different
possibilities to define a Lagrange formalism, let's consider here only
one --- the ``diagonal gauge'' Lagrange density:

\[{\cal L}_{diag}= -{1\over2} \partial^\nu A_\mu \partial_\nu A^\mu \]

For this Lagrange density, we have no problems to derive the canonical
momentum variables

\[\pi^\mu(x) = {\partial {\cal L}_{diag}\over \partial A_{\mu,t}(x)} 
=  -\partial_t A^\mu(x) \]

Canonical quantization leads to commutation relations

\[[A_\alpha(x),\pi^\beta(y)] = i \delta_\alpha^\beta \delta(x-y)\]

This defines the standard, canonical configuration space.  It does not
depend on the gravitational field, as suggested by semiclassical
postrelativistic gravity.

 \subsubsection{Particle Interpretation}

For a given tetrad field we can try to define now particle operators
similar to the scalar field separately for each component.  The only
difference is that the energy for the component 0 is negative. We can
define the vacuum state not as the state with minimal energy --- such
a state doesn't exist in the configuration space --- but the state
with maximal energy.

The configuration space now consists of four types of particles. All
of them are considered as physical in postrelativity.  For comparison,
in relativistic gauge theory, only two of them are considered as
physical.

 \subsubsection{The Incorporation of the Lorentz Condition}

One of the two additional steps of freedom is defined by the Lorentz
condition $\chi(x)=\partial_\nu A^\nu(x)$.  In classical theory, this
equation has the solution $\chi=0$.  If it is fulfilled for the
initial conditions, it will be fulfilled always.  Thus, the step of
freedom may be removed simply by making an assumption about the
initial values.

This type of incorporation of the Lorentz condition into classical
postrelativistic gauge theory has to be preferred, because the
definition of $\chi$ depends on the gravitational field, thus, this
condition should not be used to restrict the configuration space.

In the case of quantum mechanics the situation becomes more
complicate.  For non-Abelian gauge fields and Minkowski background it
is possible to define an invariant subspace with the property

 \[ \langle\phi| \chi(x) |\phi\rangle = 0. \]

For this purpose we use a splitting $\chi = \chi^+ + \chi^-$ into
adjoint operators $\chi^+$ and $\chi^-$ which allows to define the
subspace by

 \[ \chi^-(x) |\phi\rangle = 0,\; \langle\phi| \chi^+(x) = 0. \]

Here $\chi^\pm$ is the part of the operator $\chi$ consisting of
particle creation resp. destruction operators. This subspace is
invariant even if we have interaction.  Thus, the consideration can be
restricted to this subspace.  Such an invariant subspace is not
available in general.  But this does not create a conceptual problem,
because such a restriction is nice, but not necessary for
postrelativistic theory.  The assumption $\langle\chi\rangle=0$ will
be only a classical approximation.

 \subsubsection{Infinite Scattering Matrix}

For a fixed postrelativistic Minkowski background, the fixed
subdivision into space and time allows to define the subspace

\[ A_0 = 0; \; \nabla {\b A} = 0 \]

This subspace is useful for comparison with relativistic theory.  If
we consider our observation to be restricted to this subspace, we have
to make additional assumptions about the initial state to be able to
apply the theory.  In our case, we have a natural choice --- the
absence of hidden particles.  For the state after the scattering, this
condition may be not fulfilled.  We have to integrate over all
possible states of the hidden steps of freedom.

This general rule allows to make predictions about scattering of
transversal photons without being able to measure the hidden steps of
freedom.

 \subsection{Comparison With Relativistic Theory}

Comparison with relativistic theory has to be subdivided into two
parts.  At first, there is the comparison of terms which are
considered as physical in above theories, especially the scattering
matrix.  The other question is if the relativistic position to claim
non-covariant and non-gauge-invariant expressions to be non-physical
is really justified.

 \subsubsection{S-Matrix of QED}

There are different variants of relativistic QED. In the variant of
Bjorken and Drell \cite{Bjorken} the gauge condition is incorporated
into the configuration space.  Configuration space and commutation
relations of postrelativistic QED are more close to the quantization
scheme of Gupta and Bleuler \cite{Gupta} \cite{Bleuler}.

The main difference between this approach and postrelativistic QED is
the scalar product in the configuration space.  In the relativistic
variant we have an indefinite Hilbert space structure.  We obtain more
relativistic invariance in this variant, but this is obviously a
situation where postrelativism suggests to sacrifice relativistic
invariance in favour of the fundamental principles of quantum
mechanics. But this manipulation is restricted only to the gauge steps
of freedom which are considered to be unobservable in the relativistic
approach.  That's why I suppose this manipulation has no influence on
the resulting scattering matrix for the states considered as physical
by above theories.  Thus, probably the comparison of QED does not lead
to different experimental predictions.

 \subsubsection{S-Matrix of Non-Abelian Gauge Theory}

Postrelativistic gauge theory does not require a modification for the
case of non-Abelian gauge theory.  For $\chi$ we have now a more
complicate equation with interaction with the other gauge steps of
freedom:

\[ \Box \chi + [A_\mu, \partial^\mu \chi] = 0 \]

We have yet the classical solution $\chi=0$, but nonetheless in
quantum theory we cannot define an invariant subspace with
$\langle\chi\rangle=0$ as before. But the restriction of the gauge
freedom is not required in postrelativistic gauge theory.  To have an
invariant subspace is of course a nice property, but it is in no way
essential part of the theory, which is well-defined in the whole
configuration space.

The Gupta-Bleuler approach cannot be generalized straightforward to
non-Abelian gauge theory.  The restriction to the subspace $\chi=0$
leads to non-unitarity of the S-matrix.  This problem can be removed
by compensation terms known as Fadeev-Popov ghost fields
\cite{Faddeev}. Because this restriction is not required in
postrelativistic gauge theory, such a compensation is not necessary.
Thus, Faddeev-Popov ghost fields do not occur in postrelativistic
gauge theory.

This modification already leads to observable differences in the
scattering matrix.  Indeed, in relativistic gauge theory we have
(after introduction of the ghost fields) unitary evolution in the
gauge steps of freedom which are considered as physical (transversal
particles).  In postrelativistic gauge theory we have unitarity only
in the full space.  I cannot judge about the possibility to verify
this difference in real experiments, but obviously this will be much
easier compared with the case of quantum gravity.

 \subsubsection{A Quasiclassical Experiment}

The other question is if the restriction of physics to gauge-invariant
results is really justified.  Similar to the situation in
quasiclassical gravity, we have to distinguish here two notions of
gauge-invariance: c-gauge-invariance (invariance for common gauge
transformation) and q-gauge-invariance (invariance of a superposition
state for different gauge transformation on the basic states).  In
postrelativity, we have trivial c-gauge-invariance, which may be
considered as fixed by fixing the state of the vacuum to be
trivial. In relativistic gauge theory we have also q-gauge-invariance.

We can show that the concept of q-gauge-invariance leads to the same
problems as the concept of q-diffeomorphism-invariance in quantum
gravity with the classical Schr\"odinger theory limit. For this
purpose, we consider a quasiclassical experiment similar to the
experiment we have considered for gravity.  The real part of the
scalar product $\langle\psi^1|\psi^2\rangle$ defined for a pair of
solutions $(A^1,\psi^1), (A^2,\psi^2)$ is of interest here.  Without
copying this description, let's describe the results:

 \begin{itemize}

 \item In Schr\"odinger theory (multi-particle theory with Coulomb
potential for electricity), the real part of the scalar product is
observable.

 \item Postrelativistic quantum gauge theory allows to compute this
scalar product and to obtain the non-relativistic limit.

 \item The scalar product is not q-gauge-invariant. Thus, the
relativistic approach does not allow to define the scalar product.
Relativistic observable results cannot depend on this scalar product.

 \end{itemize}

The classical Maxwell equations lead to q-gauge-invariance in the
sense that they do not define the evolution of the scalar product even
classically. They have to be combined with some gauge condition.

Thus, we see, that relativistic quantum gauge theory has a problem
with the non-relativistic limit.  Of course, this is only a purely
theoretical problem.  In real QED and QFT computations the same
non-gauge-invariant terms as in postrelativity are used.  Thus, the
problem becomes obvious only if we consider the situation very
careful.

Of course, we have used here assumptions about measurability in
classical Schr\"odinger theory.  Especially we need a possibility to
create and measure a superposition.

\section{Discussion}

Last not least, let's discuss some other questions related with
postrelativity.

\subsection{Historical Context}

The harmonic coordinate equation has been often used in GR, starting
with Lanczos \cite{Lanczos} and Fock \cite{Fock}.  The Isham-Kuchar
approach \cite{Kuchar} interprets harmonic space and time coordinates
as gravity-coupled mass-less fields used to identify instants of time
and points in space.  But in the context of general relativity they
cause problems like different solutions for the same metric and
solutions which don't cover the whole solution.  Especially, a ``clock
field'' will be uncertain and measurable, different from quantum
mechanical and postrelativistic time.

Logunov et.al. have introduced the harmonic coordinate equation as a
physical equation into their {\it Relativistic Theory of Gravity}
\cite{Logunov}, \cite{Vlasov}. They have found the related
modification for the black hole and big bang scenario and the
conservation laws.  Different from postrelativity, they have
introduced a Minkowski background.  Moreover, their argumentation for
the theory was based on incorrect criticism of general relativity
\cite{Seldovich}.  This theory was the starting point for the
development of classical postrelativistic gravity.

For some of the quantization problems solved by the postrelativistic
concept, the introduction of a Newtonian background frame as a
possible solution has been recognized. For the problem of time,
theories like PG are described by Isham \cite{Isham} as ``GR forced
into a Newtonian framework''.  Isham mentions the reduction of the
symmetry group in such an approach we find in PG too.  The reason for
the rejection of this concept given there --- ``theoretical physicists
tend to want to consider all possible universes under the umbrella of
a single theoretical structure'' --- is not impressive.  The theory
defines which universes are possible.  Thus, tautologically, PG
describes all possible universes too.

Real hidden variable theories have to violate Einstein causality if
they want to predict a violation of Bell's inequality. For such
theories, it is natural to introduce a Newtonian background frame.
Bell has classified them as ``relativistic but not
Lorentz-invariant''.  A relativistic variant of the Bohm
interpretation also includes such a preferred frame.

A possible link between these two questions has been recognized
too. Isham \cite{Isham} refers to Valentini \cite{Valentini} as a
``recent suggestion that a preferred foliation of spacetime could
arise from the existence of nonlocal hidden-variables''.

Aharonov and Albert \cite{Albert} have proposed an argument against
the existence of a preferred frame in special-relativistic context,
which has been rejected by Cohen and Hiley \cite{Cohen}. Roughly
speaking, the flaw in the argument was that they have compared quantum
evolution in different Lorentz frames. But, if we adopt the preferred
frame hypothesis, the description of the evolution of the quantum
system is allowed only in the preferred frame.

Classical postrelativistic gravity and the gedanken-experiment for
quasiclassical theory have been introduced by the author 1992
\cite{Schmelzer}.

\subsubsection{Reasons for Previous Failure}

It may be asked why such a simple concept has not been tried out
before. In another formulation, it may be assumed that it has been
already tried, but has failed.  Thus, to continue the consideration of
this concept is loss of time.

Here we have to reject that non-renormalizability, which is present in
this approach, has been widely accepted as a sufficient reason to
reject a concept.  Arguments which show that this is not necessary
\cite{Ashtekar} are not very old.  Some other technical ideas like
tetrad/triad formalism (which allow to avoid non-polynomial
expressions like $\sqrt{-g}$) and functional-analytical methods for
rigorous quantization are of course necessary for rigorous
quantization of postrelativistic gravity.

Moreover, even if we assume that this approach fails, it seems
interesting to find out where it really fails, which parts of the
approach cause the failure etc.

\subsection{Esthetic Questions}

Of course, esthetic questions play an essential role in the
distinction between theories which cannot be compared directly by
experiment. Some questions we have considered --- the consideration of
the correspondence to gauge theory and the derivation of
postrelativity using first principles --- are essentially esthetic
arguments for postrelativity.  Let's consider also some other
questions:

\subsubsection{Popper's Criterion of Potential Power}

Popper \cite{Popper} has proposed a criterion of potential power. This
criterion prefers a theory which can be easier falsified as
potentially more powerful.

In this sense, already classical PG is more powerful. If PG is
correct, GR is correct too, thus, there cannot be any falsification of
GR without falsification of PG.  But, in the other direction, there
are at least theoretical possibilities to falsify PG without GR.  It
starts with the observation of nontrivial topology and the reality of
the part behind the horizon of the collapsing black hole.

If we include quantum theory into the consideration, we obtain a
really different predictive power.  As the predictions of the tree
approximation, as the prediction of the results of the quasiclassical
experiment are postrelativistic predictions, general relativity
remains silent. 

\subsubsection{What Has Been Lost?}

In discussions, a main argument against postrelativity is an
unspecified ``loss of beauty'' compared with general relativity.
Unfortunately, the opponents give usually wrong reasons, like
references to covariance or to the number of variables combined with
Occam's razor. But, of course, something really has been lost.  To
understand the issue it seems necessary to find out what has been
really lost.

At first, let's remark that it is not covariance, because any theory,
PG too, allows a covariant description.  It is also not symmetry. As
already remarked by Fock \cite{Fock}, there is no symmetry in general
relativity.

It seems useful to compare PG with an approach inside GR which defines
time as a physical clock field defined by the harmonic equation
\cite{Kuchar}.  This clock field approach remains completely inside GR
and that's why does not ``loose their beauty''.  This comparison shows
that it is also not the number of variables which makes the
difference, and it is of course not the equation used for the
variables.

\subsubsection{A Predefined Framework}

But what is it?  The only difference between the clock field approach
and PG is the metaphysical status of space and time.  In PG we have an
a-priori given framework consisting of space, time and the
affine-Galilee symmetry group.  GR --- with or without harmonic clock
fields --- lives without such a framework.  Thus, it is presence or
absence of a framework which is independent of physics which makes the
real difference.

From esthetic point of view the presence of an independent framework
can be considered as an advantage --- we obtain greater modularity.
The modular structure of the postrelativistic theory is different from
general relativity. We have clear modular parts: Hilbert space theory
--- time --- space --- configuration space --- Schr\"odinger operator.

On the other hand, there are arguments for preference of a theory
``without framework''.

First, the abstract principle ``action = reaction'' requires that the
dependence of matter from the framework leads to influence of matter
on the framework too.  We can argue that the harmonic equation $\Box t
= 0; \Box x^i = 0$ has the form of a relativistic wave equation, thus,
defines a specific, weak influence of matter on the framework which
corresponds to the specific, weak action of the framework on matter.
Nonetheless, GR is obviously a better realization of this principle.

\subsubsection{No Final Theory of Everything}

The other argument is the hope for a ``theory of everything''. It
suggests a ``unification'' of matter and framework too.  This points
to another difference between relativity and postrelativity: The
loss of hope for the final theory of everything. 

Indeed, the nontrivial physical results of general relativity are no
longer results about ``spacetime'', but results about some ether. This
defines a loss of philosophical importance of these results, and
reduces the hope that we are close to the understanding of the most
fundamental structure of the universe.  Postrelativity suggests that
there is an atomic underlying structure of the ether which has as many
rights to be considered as fundamental as the atoms of usual matter.

A unified theory of everything we know yet may be possible --- it is
known that gauge theory occurs naturally for the description of
crystal defects, by analogy it may be that particles and gauge
fields may be interpreted as different types of defects of the
crystal structure of the ether.  Nonetheless, this theory cannot
have the metaphysical status of a final theory of everything, but
simply reduces the current observation to a more fundamental level.

The experimental possibilities to observe this more fundamental
level are de-facto zero, thus, we have to give up the dream to find
the most fundamental, final theory of everything.

\subsubsection{Simplicity and Common Sense}

PG is obviously much closer to ``common sense'', that means, to the
picture of the world which is natural for our everyday experience.
This is simply shown by the ether interpretation. There is no
necessity to establish ``spacetime'', moreover ``curved spacetime''.

In this sense, PG is simpler compared with general relativity.

The fact that many expression, starting with the Einstein equations,
are essentially simpler in harmonic coordinates, also has to be
mentioned in a discussion of esthetic questions.

\subsubsection{Beauty of the Harmonic Equation}

Another criterion for beauty is the preference of a theory which
requires the existence of a certain beautiful mathematical structure,
compared with a theory which does not require this structure, if this
structure really exists, moreover if it is unique.

Applied to the harmonic equation, which is obviously required to
define the evolution of the gravitational field relative to the
background in PG, but completely unnecessary in general relativity,
this is a clear argument in favour of PG. Similarly, the existence of
the Lorentz condition is an argument for postrelativistic gauge
theory.

\subsection{The Question of Measurability}

From point of view of the relativistic paradigm, the variables we
consider as real, physical variables are not observable.  The
comparison of configuration spaces and of the quasiclassical
experiment suggest that the introduction of these steps of freedom
lead to observable consequences. Nonetheless, there seem to be no
direct measurement.

Is this a conceptual problem for postrelativity?  The answer is no.
It may be a problem of comparison of the theory with real experiments.
Indeed, if we are not able to measure some variables, we are also not
able to create the pure states of the theory in the experimental
setup.

But often this is not a real problem. Indeed, the assumption that
these steps of freedom are physical at least often allows to define a
simplest state using physical criteria for simplicity --- minimal
energy or number of particles, highest symmetry.  In these situations,
we can usually assume that we are in this simplest state. The
evolution of these states gives unique predictions also for more
complicate situations.

A nice example of this strategy is the black hole collapse. In the
initial state --- nearly Minkowski space --- we have a simplest choice
of coordinates, the affine coordinates. These initial conditions allow
to make predictions about the affine background through the collapse.

For the comparison of the predictions about a known state with
experiment there is simply no problem. The theory is and has to be
able to predict the evolution of the measurable variables of their
states.  The theory has not to be unable to predict anything else.  If
further research shows that some parts of the theory may be omitted
without observable consequences, this does not invalidate the theory.

In reality we are used to work with theories without possibility to
measure all variables.  We have no possibility to measure the color of
a quark or to observe the state of the Faddeev-Popov ghost particles,
but nonetheless use such theories successfully.

\subsection{Metaphysical Interpretation of the Background Frame}

The metaphysical interpretation of the background frame is a more or
less obvious consequence of our initial picture. The background time
describes past, present, future, and causality. As a philosophical
concept, it has to be distinguished from time measurement with clocks.
It is remarkable that the distinction between two notions of time ---
using the notions ``true time'' and ``apparent time'' --- has been
introduced already by Newton \cite{Newton}. It was already mentioned
in this definition, that they may be in principle different: ``It may
be, that there is no equable motion, whereby time may be accurately
measured. All motions may be accelerated and retarded, but the flowing
of absolute time is liable to no change.''

A similar distinction has to be made between the metaphysical concept
of position and distance. The notion ``position'' is defined by the
background space, distance by length measurement.  They become really
different in the context of a superposition of two gravitational
fields: The distance between identical positions depends on the
gravitational field. Only the existence of the notion ``position''
independent on distance measurement allows to define the scalar
product independent of the gravitational field.

\subsection{Summary}

As far as we have been able to verify, the postrelativistic principles
do not lead to serious quantization problems.  Moreover, many known
problems of the standard relativistic approach do not occur in
postrelativity:

 \begin{itemize}

 \item The problem of time, inclusive the problems related with the
Hamilton constraint in the Wheeler-DeWitt approach.

 \item Problems related with nontrivial topologies.

 \item Problems which may be related with Einstein causality, like
uncertainty of causality if the gravitational field is uncertain, the
violation of Bell's inequality, possible superluminal tunneling speed.

 \item Problems related with handling of the space diffeomorphisms,
inclusive the diffeomorphism constraints in the canonical relativistic
approach.

 \item The problem of Gribov copies in relativistic gauge theory.

 \item Problems related with the impossibility to compare different
solutions in general relativity, which is necessary for the scalar
product computation in our semiclassical experiment, the semiclassical
consideration of small modifications of the gravitational field on a
classical background.

 \item Problems related with the definition of usual observables in
quantum gravity, inclusive local energy and momentum density, which is
not observable already in classical general relativity, the vacuum
state and the number of particles which is problematic in
semiclassical general relativity, and any usual classical measurement
which becomes problematic if q-diffeomorphism-invariance is required.

 \item Problems related with the impossibility to avoid the black hole
and big bang singularities in general relativity.

 \end{itemize}

The status of the remaining known problems is not very serious from
point of view of their conceptual status.  Without diminishing the
difficulty of the technical problems, it can be said that they have
different, technical character, comparable in difficulty with the
quantization of a classical deformed crystal with an unusual nonlinear
behaviour, not conceptual problems like the problem of time.

The postrelativistic approach allows to make a lot of additional
experimental predictions in a domain where the relativistic approach
remains silent.  It predicts the evolution of variables which are
considered to be not measurable in general relativity, like time,
position, energy and momentum densities for the gravitational field,
vacuum state and number of particles in semiclassical quantum field
theory, gauge potential in postrelativistic gauge theory.  It allows
to leave the limits of semiclassical quantum gravity (tree
approximation results for gravity, superpositions of semiclassical
states).

\end{document}